\documentstyle[aps,preprint]{revtex}
\begin{document}
\title{Non-universal scaling in a model of information transmission\\
and herd behavior}
\author{Dafang Zheng,$^{1,2}$ P. M. Hui,$^{2}$ and N. F. Johnson$^{3}$}
\address{$^{1}$ Department of Applied Physics, South China University of
Technology,\\ Guangzhou 510641, P.R. China}
\address{$^{2}$ Department of Physics, The Chinese University of Hong
Kong,\\
Shatin, New Territories, Hong Kong}
\address{$^{3}$ Department of Physics, University of Oxford,\\
Clarendon Laboratory, Oxford OX1 3PU, U.K.}
\maketitle

\begin{abstract}
We present a generalized dynamical model describing the sharing of
information, and
corresponding herd behavior, in a population based on the 
recent model proposed by Egu\'iluz and Zimmermann [Phys. Rev. Lett. 
{\bf 85}, 5659 (2000)].
By introducing a size-dependent probability
for dissociation of a  cluster, we show that the exponent characterizing
the  distribution of cluster sizes  becomes
model-dependent and non-universal.
The resulting system, which provides a simplified model of a financial
market, yields power law  behavior with an easily tunable exponent.

\vspace*{0.5 true in}


\end{abstract}

\newpage
\section{Introduction}
There has been increasing interest in the study of systems
of interacting agents.  Such systems are useful for the study
of global behavior in many situations of practical importance\cite{holland}.
In the newly 
established area of econophysics\cite{stanley,bouchard},
for example, multi-agent models have been intensively
studied\cite{farmer,lux}.
Agents in a population (e.g. a market) often
do not act independently.  The collective behavior of clusters of agents,
also referred to as crowds, in which there is efficient
information and opinion sharing
among the agents, is an important factor in both real and
simulated markets\cite{us1}.
These crowds are dynamic in nature in that
there is a continual process of crowd formation and dissociation within a
competing population.

A simple model for stochastic
opinion cluster formation and information dispersal, has recently been
proposed and
studied  by Egu\'{i}luz and Zimmermann\cite{EZ}.  It is a dynamical model
(henceforth referred to as the EZ model) in which there is a continual
grouping and re-grouping of agents to form clusters.
A cluster or crowd of agents act together
(either buying or selling) and then
dissolve after the transaction has occured.
When a cluster of agents decides not to trade, i.e. an inactive state, it
may 
combine with another cluster of agents to form a bigger cluster.
Detailed numerical studies\cite{EZ}
and mean field analysis\cite{dHR1} revealed several
interesting features of the model.  For example,
it was observed that this simple model of herd behavior could
lead to a fat-tail distribution of price returns similar to that
observed in real markets.  In addition, the cluster size distribution
$n_{s}$ shows a scaling behavior of the form $n_{s} \sim s^{-5/2}$
for a range of cluster size $s$, followed by
an exponential cutoff\cite{dHR1}.
The EZ model represents a generalization
of the static percolation-type
model of Cont and Bouchard\cite{cont} in which herd formation is
described by random connection between agents, and
the cluster size distribution is found to follow the same
scaling behavior.  Several variations on the
model have been proposed and studied.  These variations include
the spreading of opinion to multiple clusters \cite{dHR1}
and inhomogeneous dissociation of clusters\cite{dHR2}.
Interestingly, it was found that the
values of the exponent characterizing the cluster size distribution
seem to be robust against these variations, i.e.
these values remain unchanged for the different variations
proposed so far.  We note that similar scaling behavior has been found
in the size distribution of businesses\cite{Takayasu,Ramsden}, although
the value of the exponent is different.

Here we introduce a generalized version of the EZ model
in which a cluster
of agents of size $s_{i}$ will dissolve
with a 
probability $s_{i}^{-\delta}$ if a transaction is made, and will combine
with
a cluster of size $s_{j}$ with a probability
$s_{i}^{-\delta}s_{j}^{-\delta}$ if a transaction is not made.
The exponent becomes {\em non-universal} and tunable with
values depending on the parameter $\delta$.
An analytic mean-field
theory is presented which provides a quantitative explanation of the
numerical
results. The EZ model is recovered as a limiting
case of our model with $\delta = 0$.

The plan of the paper is as follows.  We define our generalized model in
Sec. II and present the numerical results on the cluster size distribution.
Section III provides a mean field analysis and identifies the scaling
exponent.  The analytic results are compared with numerical simulations.
The resulting 
distribution of price-returns is presented in Sec.IV together with a
discussion on possible extensions of our model.

\section{The Model}

We consider a model with a total of $N$ agents.  Following Ref.\cite{EZ},
a cluster or crowd is a group of agents who can exchange
information efficiently and thus have a common opinion.  These agents make
the same decision at a given moment in time.  Initially, all the agents
are isolated, i.e. each agent belongs to a cluster of size unity.  As
time evolves, an agent belongs to a cluster of a certain size.  At each
timestep an agent, say the $i$-th one, is chosen at random.  Let
$s_{i}$ be the size of the cluster to which the chosen agent belongs.
Since the agents
within a cluster have a common opinion, all agents  in such a cluster tend
to imitate each
other and hence act together.  With  probability $a$ the agent, and hence
the whole
cluster, decides to  make a transaction, e.g. to buy or to sell with equal
probability.  
After the transaction, the
cluster is then broken up into isolated agents with
a probability $s_{i}^{-\delta}$, with $0 \leq \delta < 1$.  With
probability $(1-a)$ the agents decide not to make a transaction,
i.e. they wait and try to
gather more information. The other agents in the cluster follow.
In this case, another agent $j$ is chosen at random.  The two clusters
of sizes $s_{i}$ and $s_{j}$ then either combine to form a bigger cluster
with probability $s_{i}^{-\delta} s_{j}^{-\delta}$, or the two
clusters remain separate with probability $(1- s_{i}^{-\delta}
s_{j}^{-\delta})$.  Here $a$ can be treated as a parameter reflecting
the investment rate showing how frequent a transaction is made.  Our
model thus represents a generalization of the basic EZ model to the case
in which a cluster of agents may stay together to form a group
{\em after} making a transaction.
The probability of dissociation $s_{i}^{-\delta}$ implies
that larger clusters
have a larger tendency to remain grouped while smaller
clusters are easier
to break up\cite{dHR3}.  For the special case of $\delta = 0$, our model
reduces 
to the EZ model. 

In the EZ model ($\delta = 0$), clusters of agents break up  after a
transaction.  Here, our model includes a dissociation  probability depending
on the
cluster size - this feature may be invoked to mimic practical aspects of a
financial
market, such as the effect of news arrival. Imagine one of
the agents in a cluster of size $s_{i}$ receives some external news
with probability $a$ at a given timestep.  This external news suggests
that he, and hence the other members of his cluster, should immediately
trade
(buy or sell).  Since the news  is external, the crowd act together in this
one moment,
leaving the cluster with a finite probability of subsequently dissociating.
Suppose that they sense, e.g. from the resulting price-movement, that they
are a member of a large crowd of like-minded agents: in practice many
traders like to
feel part of a larger crowd for reassurance. We therefore assume that the
crowd breaks up
with a size-dependent probability $p(s_{i})$, where $p(s_{i}) = 1$ for
$s_{i}=1$ and
$p(s_{i})$  decreases monotonically as $s_{i}$ increases.  By contrast, with
probability
($1-a$) there  is no news arrival from outside.  The agent in the chosen
cluster,
uncertain about whether to buy or sell, makes contact with an agent in
another cluster of 
size
$s_{j}$.  The agents share information and come up with a new opinion.  Each
of them then
separately tries to persuade the other members of his  cluster of the new
opinion.  With
probability
$p(s_{i})$ ($p(s_{j})$)  the opinion of cluster $i$ ($j$) changes to the new
opinion. 
Thus,  the two clusters combine with probability $p(s_{i})p(s_{j})$.  It
turns 
out that this particular form of the two combined modifications to the EZ
model, can be treated using our mean field analysis.  As a specific example,
our
numerical  simulations are carried out for the case in which $p(s) \sim
s^{-\delta}$.

Let $n_{s}$ be the number of clusters of size $s$.  Figure 1 shows the
results of
numerical  simulations on the cluster size distribution in the steady state,
for various values of the parameter $\delta$.  The results are obtained for
a system with
$N=10^{4}$ and 
$a=0.3$ \cite{EZ}.  Averages are taken over a time window of $10^{6}$
time steps 
after the transient behavior has disappeared, together with a configuration
average over 100 different runs with different initial conditions.
The $\delta = 0$ results
give the features in the EZ model.  For a range of $s$, $n_{s} \sim
s^{-\beta}$ with $\beta = 5/2$ \cite{EZ}. Deviation from the scaling
behavior sets in at a value of $s$ depending on the value of the
parameter $a$.  These features are consistent with previous
numerical\cite{EZ}
and analytical studies\cite{dHR1}.
For $0 \leq \delta < 1$, it is observed that
the size distribution $n_{s}$ still scales with $s$ in a range of $s$
as in the EZ model.  However, the exponent becomes model dependent
and hence {\em non-universal}.  The data shows that the exponent
$\delta$ is consistent with the behavior $n_{s} \sim s^{-\beta(\delta)}$,
where $\beta(\delta) = 5/2 - \delta$.  A mean field analysis can be
used to extract this scaling behavior, as will be described in the
next section.

It is interesting to note that several attempts have been made to modify
the EZ model.  These extensions include, for
example, the study by d'Hulst and
Rodgers on democrazy versus dictatorship by incorporating an
inhomogeneous investment rate in the population\cite{dHR2}
and also allowing 
rumor to spread to multiple clusters in one time step after a
chosen cluster decides not to make a transaction\cite{dHR1}.  All the
extensions 
proposed so far give $\beta = 5/2$, hence the value seems to be robust.
The present model incorporates a size-dependent dissociation probability of
a 
cluster after a transaction and leads to a tunable and model-dependent
$\beta(\delta)$.  Thus our model actually gives a set of models
with different values of $\beta$, similar to the case of changing
a system from one universality class to another in problems in critical
phenomena.  In fact, the situation is reminiscent of the non-universal
exponent of conductivity in continuum percolation\cite{Halperin,hui1}.
In percolation problems\cite{stauffer1},
it is known that the effective
conductivity for a system consisting of insulators and conductors
exhibits the scaling behavior $\sigma_{e} \sim (p-p_{c})^{t}$ near
the percolation threshold $p_{c}$.  The exponent $t$ is universal
in that its value depends only on the dimension of the system, regardless
of other details, e.g. lattice type.  However,
if the conductances $\sigma$ of the
conductors follow a distribution of the form
$P(\sigma) \sim \sigma^{-\delta}$
with $0 < \sigma < 1$, the $t$-exponent\cite{Halperin}
and other related 
properties\cite{hui1,hui2} become non-universal with
exponents taking on a value depending on $\delta$.
It should be noted that it is not so surprising to see a connection
between percolation and model for herd behavior.
In the model of Cont and Bouchard\cite{cont}, the EZ model\cite{EZ}
and their variations\cite{stauffer2},
an agent could be connected to any one of $(N-1)$ other agents to form a
cluster.  These models hence represent a problem of connectivity in high
dimensions.  Several other percolation type models\cite{stauffer3,stauffer4}
have also been proposed to explain the phenomena observed in real markets.

\section{Mean field analysis}

The cluster size distribution in the EZ model can be studied via a mean
field analysis\cite{dHR1}.  The treatment can be extended to the present
model to extract the scaling behavior of $n_{s}$, though the algebra
is more complicated.  Let $n_{s}(t)$ be the number of
clusters of size $s$ at time $t$.  At a certain time, $n_{s}(t)$
changes as a result of the collective action
of the members
of the cluster containing the chosen agent.
A master equation can thus
be written down: 
\begin{equation}
N \frac{\partial n_{s}}{\partial t} =
-a s^{1-\delta} n_{s} + \frac{(1-a)}{N} \sum_{r=1}^{s-1}
r^{1-\delta} n_{r} (s-r)^{1-\delta} n_{s-r}
- \frac{2(1-a)s^{1-\delta} n_{s}}{N} \sum_{r=1}^{\infty} r^{1-\delta}n_{r}
\end{equation}
for $s > 1$.  Each of the terms on the right hand side of Eq.(1)
represents the consequence of a possible action of the agent.  The first
term describes the dissociation of a cluster of size $s$ after a
transaction is made.  The second term represents coagulation of two clusters
to form a cluster of size $s$.  The third term represents the
combination of a cluster of size $s$ with another cluster.  For clusters
of size unity ($s=1$), we have
\begin{equation}
N \frac{\partial n_{1}}{\partial t} =
a \sum_{r=2}^{\infty} r^{2-\delta}n_{r}
- \frac{2 (1-a)n_{1}}{N} \sum_{r=1}^{\infty} r^{1-\delta}n_{r}.
\end{equation}
Here, the first term comes from the dissociation of any clusters
into isolated agents and the second term describes the combination of a
cluster of size unity with another cluster.  In the steady state,
$\frac{\partial n_{s}}{\partial t} = 0$, we have
\begin{equation}
s^{1-\delta} n_{s} = A \sum_{r=1}^{x-1} r^{1-\delta}
(s-r)^{1-\delta} n_{r} n_{s-r}
\end{equation}
for $s>1$, and 
\begin{equation}
n_{1} = B \sum_{r=2}^{\infty} r^{2-\delta} n_{s}
\end{equation}
for $s =1$, where 
\begin{equation}
A = \frac{1-a}{Na + 2(1-a)\sum_{r=1}^{\infty} r^{1-\delta}n_{r}},
\end{equation}
and 
\begin{equation}
B = \frac{Na}{2(1-a) \sum_{r=1}^{\infty} r^{1-\delta} n_{r}}.
\end{equation}

The aim here is to extract the scaling behavior.  Invoking a generating
function approach\cite{wallace}, we let
\begin{equation}
G(\omega) = \sum_{r=0}^{\infty} r^{1-\delta} n_{r} e^{-\omega r} = g(\omega)
+ 
n_{1} e^{-\omega}, 
\end{equation}
where $g(\omega) = \sum_{r=2}^{\infty} r^{1-\delta} n_{r} e^{-\omega r}$.
It follows from Eq.(3) that the function $g(\omega)$ satisfies
the equation 
\begin{equation}
g^{2}(\omega) + (2n_{1} e^{-\omega} - \frac{1}{A}) g(\omega)
+ n_{1}^{2} e^{-2\omega} = 0.
\end{equation}
Note that $A$ can be expressed in terms of $n_{1}$ and $g(0)$, and
\begin{equation}
g(\omega) = \frac{1}{4A} (1 - \sqrt{1 - 4n_{1} A e^{-\omega}})^{2}.
\end{equation}
The number of clusters of size $s$ can be found formally by
\begin{equation}
n_{s} = \frac{1}{s^{1-\delta} s!} \frac{\partial^{s}G}{\partial
z^{s}}|_{z=0}, 
\end{equation}
where $z = e^{-\omega}$.  The resulting expression for $n_{s}$ is
\begin{equation}
n_{s} = \frac{(2s-2)! (1-a)^{s-1} (\sum_{r=1}^{\infty}
r^{1-\delta}n_{r})^{s}
[(1-a) \sum_{r=1}^{\infty} r^{1-\delta} n_{r} + Na]^{s}}
{(s!)^{2} s^{-\delta} [Na + 2(1-a)\sum_{r=1}^{\infty} r^{1-\delta}n_{r}]
^{2s-1}}. \
\end{equation}
Invoking Sterling's formula yields
\begin{equation} 
n_{s} \sim N \left[ \frac{ 4(1-a)[(1-a) + \frac{Na}{\sum_{r=1}^{\infty}
r^{1-\delta}n_{r}}]}
{[\frac{Na}{\sum_{r=1}^{\infty} r^{1-\delta}n_{r}} + 2(1-a)]^{2}}
\right]^{s} 
s^{-(\frac{5}{2} - \delta)}.
\end{equation}
For $\delta = 0$, the sum $\sum_{r=1}^{\infty} r^{1-\delta} n_{r} = N$
and the previous results of Refs.\cite{EZ,dHR1} are recovered.
For $\delta \neq 0$,
it is difficult to solve for $n_{s}$.  Since the summations in Eqs.(11)
and (12) give a number, our result shows that
$n_{s} \sim s^{-\beta(\delta)}$ with $\beta(\delta) = 5/2 - \delta$
for the present model.  This $\delta$-dependent exponent is also
indicated in Fig. 1 (lines are a guide to the eye).
We note that the scaling behavior is masked
by the behavior of the term in the squared brackets in Eq.(12) for
large values of $s$, similar to the situation for the EZ model\cite{dHR1}.

\section{Discussion}

Egu\'iluz and Zimmermann\cite{EZ} applied their model to study the
distribution 
of price returns.  A price can be generated according to
\begin{equation} 
P(t+1) = P(t) \exp(s'/\lambda),
\end{equation} 
where $\lambda$ is a parameter for the liquidity of the market.  The
price return $R(t) = \ln P(t) - \ln P(t-1)$ is defined to be
the relative number of agents buying or selling at a time with
$s' = s$ for a cluster of agents deciding to buy, and $s' = -s$
for a cluster deciding to sell at a given timestep.  Numerical results for
the EZ model showed that the distribution of returns $P(R) \sim
R^{-\alpha}$ with $\alpha = 3/2$.  We have carried out similar
calculations for our model.  Figure 2 shows the price return
distributions for different values of $\delta$ on a log-log scales.
As for the cluster size distributon, the exponent $\alpha$ is now
{\em non-universal} and takes on the value $3/2 - \delta$, which is
also the value of $\beta(\delta) -1$ \cite{EZ}.

In summary, we have proposed and studied the cluster size distribution,
and the price return distribution, of a generalized version of the EZ model.
Our model is a dynamical model for herd behavior and information sharing in
a 
population.  By introducing a probability for dissociation of a
cluster depending on its size, the exponent characterizing the
cluster size distribution takes on a model-dependent non-universal
value.  Our model thus provides a simple way for tuning the power law
behavior.  Several extensions are immediately possible.  Our particular
choice of the form of the probability for dissocation of clusters
allows us to tune the exponent within a range of unity.  Changing
the functional form of the probability slightly may alter the range of
values
in which the exponent can be tuned.
A positive and negative value of $\alpha$ in our model
may lead to different features in the cluster size distribution.
Our model can 
also be extended to study the size distribution of businesses. A model
similar to the EZ
model has already been proposed in this context\cite{dHR3}, however  the
scaling behavior
seems to be non-universal for data from different
countries\cite{Takayasu,Ramsden}.  The
present model thus provides a possible extension to cope with this observed
non-universality. 

\begin{center}
{\bf ACKNOWLEDGMENTS}
\end{center}

One of us (DFZ) acknowledges the 
support from the Natural Science Foundation of Guangdong Province, China.
The visit of DFZ to the Department of Physics at CUHK,  
during which this work was initiated, was 
supported in part by a Grant (CUHK4129/98P)
from the Research Grants Council of the Hong Kong SAR Government.

\newpage \centerline{\bf FIGURE CAPTIONS}

\bigskip
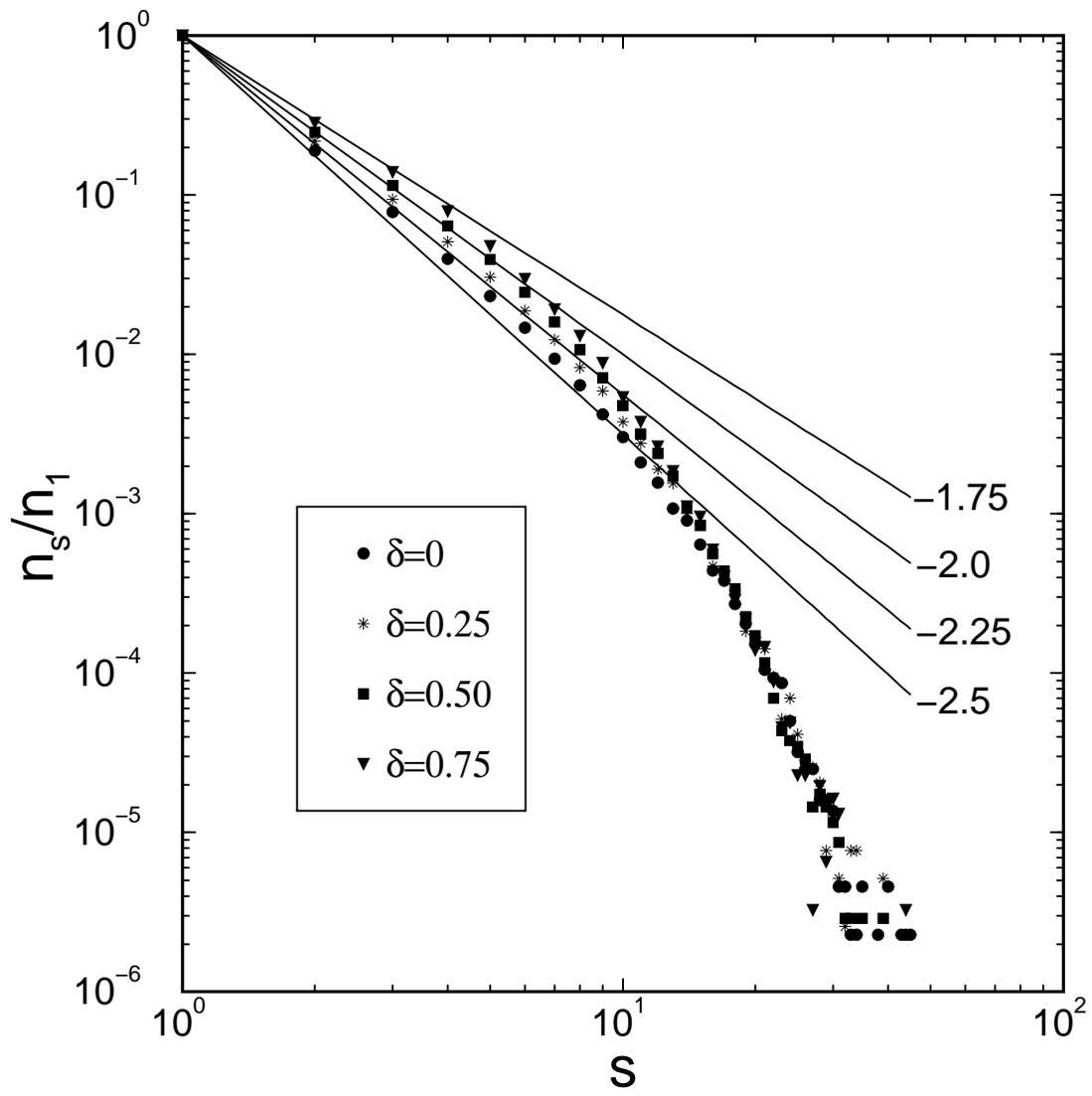
\noindent Figure 1: The size distribution $n_{s}/n_{1}$ as a function
of the size $s$ on a log-log scale for different values of
$\delta$ obtained by numerical simulations (symbols).  The values of
$\delta$ used in the calculations are: $\delta = 0, 0.25, 0.50, 0.75$.
The solid lines are a guide to the eye corresponding to exponents
$\beta = -2.5, -2.25, -2.0, -1.75$ respectively.

\bigskip
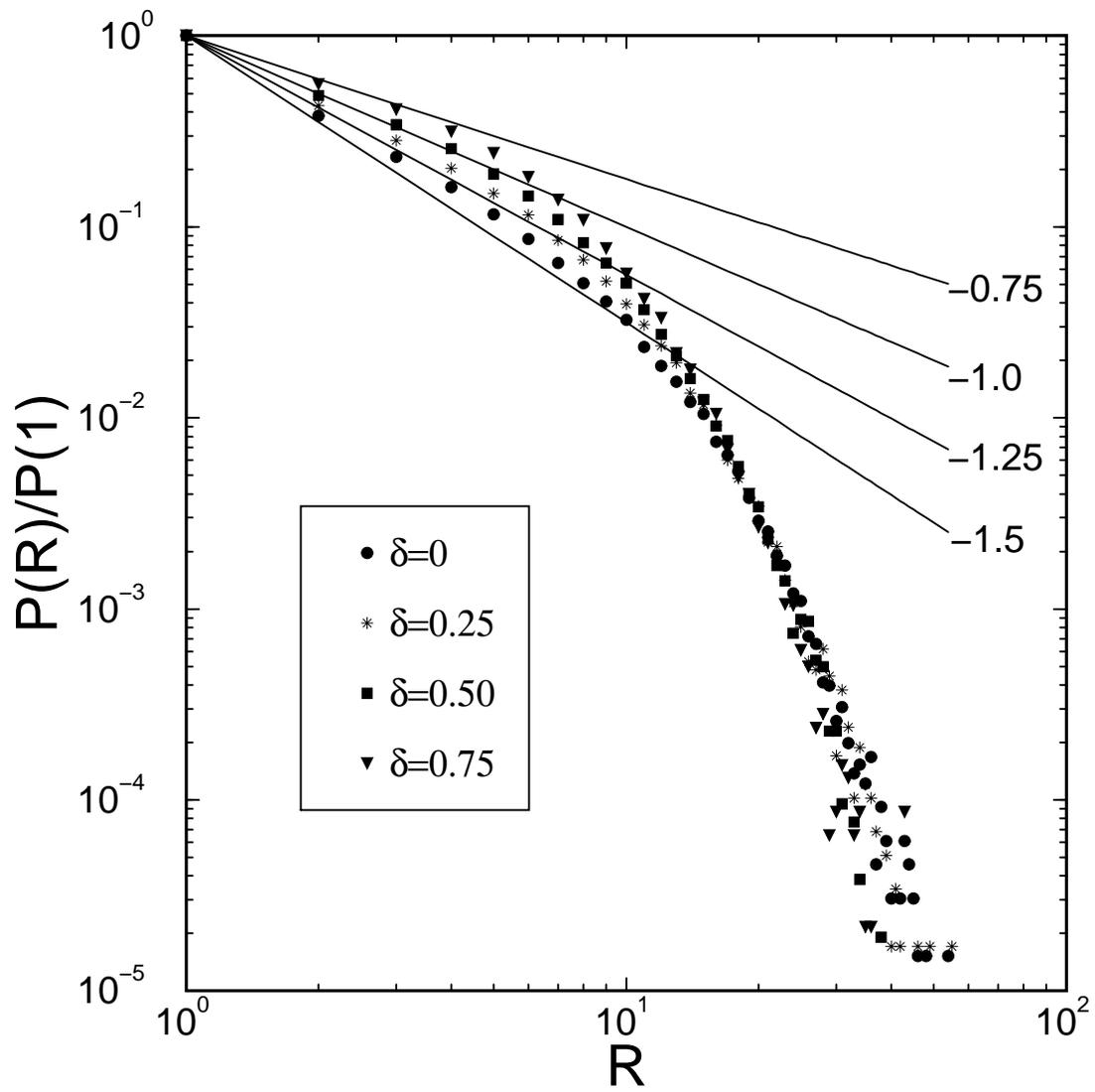
\noindent Figure 2:  The distribution of price returns $P(R)/P(1)$
as a function of $R$ on a log-log scale for different values of
$\delta$ (symbols).  The values of $\delta$ used in the calculations are:
$\delta = 0, 0.25, 0.50, 0.75$.  
The solid lines are a guide to the eye corresponding
to exponents $\alpha = -1.5, -1.25, -1.0, -0.75$ respectively.

\end{document}